\documentclass[twocolumn,preprintnumbers,showpacs,pre,floatfix,amsmath,amssymb,notitlepage]{revtex4}
\usepackage{epsfig}
\usepackage{subfigure}
\usepackage{amsmath}
\usepackage{color}
\usepackage{amssymb}
\usepackage{setspace}
\usepackage{graphicx}
\usepackage{dcolumn}
\usepackage{bm}
\usepackage{times}
\usepackage{enumerate}
\usepackage{footnote}
\input epsf

\graphicspath{{figures/}}

\begin{document}

\author{Per Sebastian Skardal}
\email{skardal@colorado.edu} 
\affiliation{Department of Applied Mathematics, University of Colorado at Boulder, Colorado 80309, USA}

\author{Juan G. Restrepo} 
\affiliation{Department of Applied Mathematics, University of Colorado at Boulder, Colorado 80309, USA}

\title{Hierarchical Synchrony of Phase Oscillators in Modular Networks}


\begin{abstract}
We study synchronization of sinusoidally coupled phase oscillators on networks with modular structure and a large number of oscillators in each community. Of particular interest is the hierarchy of local and global synchrony, i.e., synchrony within and between communities, respectively. Using the recent ansatz of Ott and Antonsen, we find that the degree of local synchrony can be determined from a set of coupled low-dimensional equations. If the number of communities in the network is large, a low-dimensional description of global synchrony can be also found. Using these results, we study bifurcations between different types of synchrony. We find that, depending on the relative strength of local and global coupling, the transition to synchrony in the network can be mediated by local or global effects.
\end{abstract}


\maketitle

\section{Introduction}

Large networks of coupled oscillators are pervasive in science and nature and serve as an important model for studying emergent collective behavior. Some examples include synchronized flashing of fireflies \cite{FireFly}, cardiac pacemaker cells \cite{Pacemaker}, walker-induced oscillations of some pedestrian bridges \cite{Millenium}, Josephson junction circuits \cite{Marvel1}, and circadian rhythms in mammals \cite{Circadian}. A paradigmatic model of the emergence of synchrony in systems of coupled oscillators is the Kuramoto model \cite{Kuramoto1}, in which each oscillator is described by a phase angle $\theta_n$ that evolves as
\begin{equation}\label{eqH}
\dot \theta_n = \omega_n + \frac{1}{N}\sum A_{nm} \sin(\theta_m - \theta_n),
\end{equation}
where $\omega_n$ is the intrinsic frequency of oscillator $n$, $A_{nm}$ represents the strength of the coupling from oscillator $m$ to oscillator $n$, and $n,m=1,\dots,N$. The classical all-to-all Kuramoto model corresponds to $A_{nm} = k$. The study of generalizations of the Kuramoto model has become an important area of research. Some examples of such generalizations include systems with time-delays \cite{Lee1}, network structure \cite{Restrepo1,Barlev1}, non-local coupling \cite{Martens1}, external forcing \cite{Childs1}, non-sinusoidal coupling \cite{Daido1},  cluster synchrony \cite{Skardal1}, coupled excitable oscillators \cite{Alonso1}, bimodal distributions of oscillator frequencies \cite{Martens2}, phase resetting \cite{Levajic1}, time-dependent connectivity \cite{So1}, noise \cite{Nagai1}, and communities of coupled oscillators \cite{Pikovsky1,Barreto1,Montbrio1,Kawamura1}.

In this paper we study the case where the coupling strength is not uniform, but rather defines a network that has strong modular, or community, structure. Synchrony on heterogeneous networks has been studied in the past, both for phase oscillator systems \cite{Restrepo1} and other dynamical systems \cite{Pecora1}. Much recent work has focused on the synchronization of phase oscillators on networks with modular structure \cite{Pikovsky1,Barreto1,Montbrio1,Kawamura1}. While the link between community topology and synchronization is well established~\cite{Arenas1}, there are few analytical results that describe synchronization in modular networks. Reference~\cite{Pikovsky1} developed a framework to study a general number of communities, assuming that oscillators within communities are identical. Reference \cite{Barreto1} analyzed the linear stability of the incoherent state for a system of coupled communities of heterogeneous phase oscillators. The same system was considered in Ref.~\cite{OA1}, where a set of coupled low-dimensional equations governing the dynamics of the community order parameters was formulated. Here, we study this system of equations, finding for some important cases analytical expressions for local and global order parameters describing synchronization within communities and on the whole network, respectively.  We find that, in the limit of a large number of communities, the Ott-Antonsen ansatz introduced in Ref.~\cite{OA1} can be used to obtain a low dimensional description of community synchrony. Using this description, we characterize the phase space of the system where the parameters are the local and global coupling. One of our results is that, depending on the relative strength of local and global coupling, the transition to synchrony in the network can be mediated by local or global effects.

This paper is organized as follows. In Sec. II we describe the model. In Secs. III and IV we present in detail the local and global dimensionality reductions, respectively. In Sec. V we discuss the effect of community structure of the network  on the dynamics and how it promotes hierarchical synchrony. In Sec. VI we discuss how our results generalize when certain heterogeneities are introduced into the network. In Sec. VII we conclude this paper by discussing our results.

\section{Model description}

We are interested in studying coupled oscillators on a network with strong community structure such that (i) the coupling strength between oscillators within the same community is much larger than the coupling strength between oscillators in different communities and (ii) the intrinsic frequency for an oscillator is drawn from a distribution specific to the community to which that oscillator belongs. Condition (i) serves as a model of situations where all the coupling strengths have similar magnitude, but the density of connections between communities is less than the density of connections within a community. The motivation for condition (ii) is that oscillators in different communities could have different frequency distributions due to different functional needs (e.g., as in cardiac myocytes in different regions of the heart \cite{mathphys}), or as an approximation to fluctuations inherent to large but finite communities.  Thus, for a network with $C$ communities labeled $\sigma=1,2,\dots,C$ where community $\sigma$ contains $N_\sigma$ oscillators, we assume that the coupling matrix $A$ in Eq.~(\ref{eqH}) can be written in block form as $A_{nm}=K^{\sigma\sigma'}$, where $\sigma$ and $\sigma'$, respectively, denote the communities to which oscillators $n$ and $m$ belong. Furthermore, we assume that the intrinsic frequencies for oscillators in community $\sigma$ are drawn from a distribution particular to that community, denoted by $g_\sigma(\omega)$. We denote the fraction of oscillators in community $\sigma$ by $\eta_\sigma=N_\sigma/N$, where $N$ is the total number of oscillators in the whole network.

With this notation, Eq.~(\ref{eqH}) results in the following system, considered in Refs.~\cite{Barreto1,OA1}:
\begin{align}\label{eqModel}
\dot{\theta}_n^\sigma=\omega_n^\sigma + \sum_{\sigma'=1}^C\eta_\sigma\frac{K^{\sigma\sigma'}}{N_{\sigma'}}\sum_{m=1}^{N_{\sigma'}}\sin(\theta_m^{\sigma'}-\theta_n^\sigma),
\end{align}
where $\theta_n^\sigma$ denotes the phase of an oscillator in community $\sigma$, $\sigma=1,\dots C$, $n=1,\dots,N_\sigma$, and the intrinsic frequency $\omega_n^\sigma$ is randomly drawn from the distribution $g_\sigma(\omega)$. Next, in order to measure synchrony within and between communities we define the local and global order parameters
\begin{align}
z_{\sigma}&=r_\sigma e^{i\psi_\sigma}=\frac{1}{N_\sigma}\sum_{m=1}^{N_\sigma}e^{i\theta_m^\sigma}, \label{eqOrdr} \\ 
 Z&=Re^{i\Psi}= \sum_{\sigma=1}^C \eta_\sigma z_\sigma, \label{eqOrdR}
\end{align}
respectively, such that $r_\sigma$ measures the degree of local synchrony in community $\sigma$ and $R$ measures the degree of global synchrony over the entire network. We note that the linear stability of the incoherent state in this model was studied in Ref.~\cite{Barreto1} (see also \cite{Montbrio1}). 

\section{Local dimensionality reduction}

In this section, we will study local synchrony by assuming there are a large number of oscillators $N_\sigma$ in each community. Using the definition of $z_\sigma$ in Eq.~(\ref{eqOrdr}), we simplify Eq.~(\ref{eqModel}) to
\begin{align}
\dot{\theta}_n^\sigma=\omega_n^\sigma+\frac{1}{2i}\sum_{\sigma'=1}^C \eta_{\sigma'} K^{\sigma\sigma'}(z_{\sigma'}e^{-i\theta_n^\sigma}-z_{\sigma'}^*e^{i\theta_n^\sigma}),
\end{align}
where $^*$ denotes complex conjugate. We now move to a continuum description by taking the limit $N,N_\sigma\to\infty$ in such a way that all $\eta_\sigma$ remain constant. Accordingly, we introduce the density function $f_\sigma(\theta,\omega,t)$ that represents the density of oscillators in community $\sigma$ with phase $\theta$ and natural frequency $\omega$ at time $t$. Since the number of oscillators in each community is conserved, $f_\sigma$ satisfies the local continuity equation, $\partial_t f_\sigma + \partial_{\theta^\sigma}(f_\sigma\dot{\theta}^\sigma)=0$, or
\begin{align}\label{eqContIntra}
\partial_t f_\sigma + \partial_{\theta^\sigma} \left\{f_\sigma \left[\omega^\sigma + \sum_{\sigma'=1}^C \eta_{\sigma'} K^{\sigma\sigma'}\text{Im}(z_{\sigma'}e^{-i\theta^\sigma}) \right]\right\}=0.
\end{align}

Following Ott and Antonsen \cite{OA1}, we expand $f_\sigma(\theta,\omega,t)$ in a Fourier series, $f_\sigma(\theta,\omega,t)=\frac{g_\sigma(\omega)}{2\pi}\left(1+\sum_{n=1}^\infty \widehat{f}_{\sigma,n}(\omega,t)e^{in\theta} + c.c.\right)$, and make the ansatz $\widehat{f}_{\sigma,n}(\omega,t)=a_\sigma^n(\omega,t)$, namely
\begin{align}\label{eqAnsIntra}
f_\sigma(\theta,\omega,t)=\frac{g_\sigma(\omega)}{2\pi}\left(1+\sum_{n=1}^\infty a_\sigma^n(\omega,t)e^{in\theta} + c.c.\right),
\end{align}
which, when introduced in Eq.~(\ref{eqContIntra}), yields a single ordinary differential equation (ODE)
\begin{align}\label{eqODEa}
\dot{a}_\sigma+i\omega a_\sigma+\frac{1}{2}\sum_{\sigma'=1}^C\eta_{\sigma'}K^{\sigma\sigma'}(z_{\sigma'}a^2_\sigma-z_{\sigma'}^*)=0,
\end{align}
where $z_\sigma$ in the continuum limit is given by
\begin{align}
z_{\sigma}&=\int_{-\infty}^\infty\int_0^{2\pi}f_{\sigma}(\theta,\omega,t)e^{i\theta}d\theta d\omega\nonumber\\
&= \int_{-\infty}^\infty g_{\sigma}(\omega)a^*_{\sigma}(\omega,t)d\omega.
\end{align}


Finally, by letting the distribution of frequencies $g_\sigma$ be a Lorentzian with spread $\delta_\sigma$ and mean $\Omega_\sigma$, i.e. $g_\sigma(\omega)=\delta_\sigma/\{\pi[\delta_\sigma^2+(\omega-\Omega_\sigma)^2]\}$, we can calculate $z_{\sigma}$ by closing the $\omega$ contour of integration with the lower-half semicircle of infinite radius in the complex plane and evaluating $a_{\sigma}^*(\omega,t)$ at the enclosed pole of $g_{\sigma}$:
\begin{align}\label{eqIntIntra}
z_{\sigma}=a_{\sigma}^*(\Omega_{\sigma}-i\delta_{\sigma},t).
\end{align}
Thus, by evaluating Eq.~(\ref{eqODEa}) at $\omega=\Omega_{\sigma}-i\delta_{\sigma}$, we close the dynamics for $z_\sigma$:
\begin{align}\label{eqSimpIntra}
\dot{z}_\sigma + (\delta_\sigma-i\Omega_\sigma)z_\sigma + \frac{1}{2}\sum_{\sigma'=1}^C\eta_{\sigma'}K^{\sigma\sigma'}(z_{\sigma'}^*z_{\sigma}^2-z_{\sigma'})=0,
\end{align}
which defines $C$ complex ODEs, or equivalently $2C$ real ODEs, given by
\begin{align}
\dot{r}_\sigma &= -\delta_\sigma r_\sigma + \frac{1-r_\sigma^2}{2}\sum_{\sigma'=1}^C \eta_{\sigma'}K^{\sigma\sigma'}\text{Re}(z_{\sigma'}e^{-i\psi_\sigma}),\label{eqrho1}\\
\dot{\psi}_\sigma &= \Omega_\sigma+\frac{r_\sigma^2+1}{2r_\sigma}\sum_{\sigma'=1}^C \eta_{\sigma'}K^{\sigma\sigma'}\text{Im}(z_{\sigma'}e^{-i\psi_\sigma}).\label{eqpsi1}
\end{align}

Equation~(\ref{eqSimpIntra}) was formulated originally in Ref.~\cite{OA1}, but its consequences for hierarchical synchrony have not been studied in detail. Equations~(\ref{eqrho1}) and (\ref{eqpsi1}) describe the dynamics of local synchrony. The synchrony of community $\sigma$ is described by the magnitude of its order parameter $r_{\sigma}$ and phase $\psi_{\sigma}$. The phase variable $\psi_\sigma$ obeys an equation similar to that of the network-coupled Kuramoto model, Eq.~(\ref{eqH}), but the effect of community $\sigma'$ on community $\sigma$ is modulated by the degree of synchrony of community $\sigma'$, $r_{\sigma'}$, and its relative size $\eta_{\sigma'}$. In contrast to the Kuramoto model, each community has an additional variable $r_{\sigma}$ which evolves in conjunction with the phase variable $\psi_\sigma$. In this sense, the dynamics of the community order parameters resembles a network of coupled complex Ginzburg-Landau oscillators \cite{Hakim1}. 
 
 \begin{table}[b]
\begin{tabular}{|c|c|}
\hline
State Variables & Description \\
\hline
$r_\sigma$ & degree of local synchrony of community $\sigma$ \\
$R$ & degree of global synchrony \\
$\psi_\sigma$ & phase of local order parameter $\sigma$ \\
$\Psi$ & phase of global order parameter \\
\hline
Parameters & Description \\
 \hline
 $k$ & local coupling strength  \\
 $K$ & global coupling strength \\
 $\delta$ & local frequency spread \\
 $\Delta$ & global frequency spread \\
 $\Omega_\sigma$ & mean intrinsic frequency of community $\sigma$ \\
 $N_\sigma$ & size of community $\sigma$ \\
 $C$ & total number of communities \\
 \hline
\end{tabular}
\setlength{\abovecaptionskip}{10pt}
 \caption{Summary of local and global state variables and parameters of the system.}
 \label{table}
\end{table}
 
 In what follows, we will consider the illustrative case in which all communities have the same size and spread in natural frequencies, i.e. $\eta_\sigma=C^{-1}$ and $\delta_\sigma=\delta$. Furthermore, we let the coupling strength within each community be the same, as well as the coupling strength between oscillators in different communities. We assume the coupling strength within communities is much larger than that between communities, namely
 \begin{equation}\label{ksigma}
 K^{\sigma\sigma'} = \left\{\begin{array}{ll}Ck & \text{ if }\sigma=\sigma'\\ K &\text{ otherwise,}\end{array}\right.
 \end{equation}
where $k$ and $K$ are of the same order. We clarify that the local coupling strength $Ck$ is chosen so that the local coupling within a community is of the same order as the sum of the coupling to every other community. More generally, a local coupling strength of the form $K^{\sigma\sigma}=k/\epsilon$ with $\epsilon\ll1$ can be analyzed from our results by rescaling $k$ by a factor of $C\epsilon$. In section VI we relax these assumptions and discuss the case where community sizes, spread in frequency distributions, and coupling strengths vary from community to community. We now use the definition of $Z$ in Eq.~(\ref{eqOrdR}) to rewrite the system in Eqs.~(\ref{eqrho1}) and (\ref{eqpsi1}) as
\begin{align}
\dot{r}_\sigma & = -\delta r_\sigma  + \left(k-\frac{K}{C}\right)r_\sigma \frac{1-r_\sigma^2}{2}\nonumber \\&\hskip2ex+K\frac{1-r_\sigma^2}{2} R \cos(\Psi-\psi_\sigma),\label{eqrho}\\
\dot{\psi}_\sigma &=\Omega_\sigma +K\left(\frac{r_\sigma^2+1}{2 r_\sigma}\right)R\sin(\Psi-\psi_\sigma).\label{eqpsi}
\end{align}
We note that although we will let $C\to\infty$ in the next section, Eqs.~(\ref{eqrho}) and (\ref{eqpsi}) are valid when $C$ is any positive integer and can be used to study synchrony on networks with a small number of communities.

Finally, we assume that the mean frequencies $\Omega_\sigma$ are drawn from a distribution $G(\Omega)$, which we assume to be Lorentzian with spread $\Delta$ and mean $\Gamma$. However, by entering a rotating frame, we can set $\Gamma=0$ without any loss of generality. For the sake of convenience we summarize all local and global state variables and parameters of the system in Table \ref{table}. We note that choosing a Lorentzian distribution for $G(\omega)$ is a natural choice if the heterogeneity in the distributions $g_{\sigma}(\omega)$ is assumed to originate from fluctuations arising from the random sampling of frequencies from the same Lorentzian distribution. In this case, since a sum of Lorentzian random variables has a Lorentzian distribution, the distribution of the average frequencies in finite communities is Lorentzian.

\begin{figure}[t]
\centering
\epsfig{file =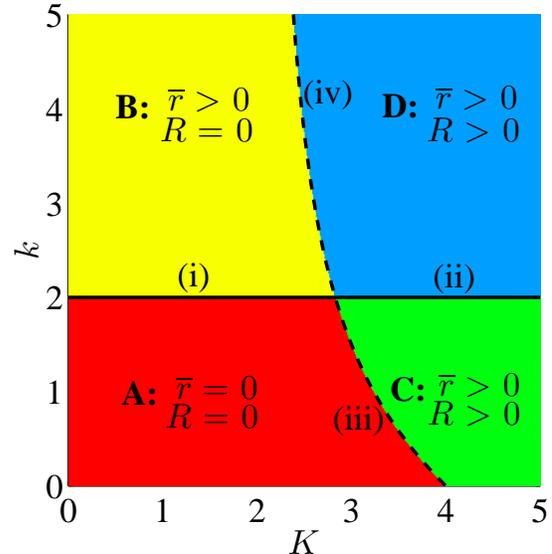, clip =,width=0.94\linewidth }
\caption{(Color online) Bifurcation diagram in $(K,k)$ parameter space for Eq.~(\ref{eqModel}) with $\delta=\Delta=1$. Regions A, B, C, and D (described in the text) are denoted in red, yellow, green, and blue, respectively, with bifurcations (i)-(iv) indicated by solid and dashed curves.} \label{BifD}
\end{figure}

Before analyzing Eqs.~(\ref{eqrho}) and (\ref{eqpsi}), we illustrate the behavior of the local and global order parameters $\delta=\Delta=1$ over a range of values for $K$ and $k$. We define $\overline{r}=C^{-1}\sum_{\sigma=1}^C r_\sigma$ as a measure of local synchrony and show the behavior  of $\overline{r}$ and $R$ in Fig.~\ref{BifD}. While this behavior will be deduced from the analysis that follows, we find it convenient to present the phase space now to provide a framework for our subsequent analyses. We note that although the diagram above is theoretical, we present plots of $R$ and $\bar r$ following various paths in the diagram, and all show excellent agreement with the theory. In the parameter space $(K,k)$, we find the following four regions: region A where $\overline{r},R=0$ (bottom left red), region B where $\overline{r}>0,R=0$ (top left yellow), and regions C and D where $\overline{r},R>0$ (bottom right green and top right blue, respectively). In region A there is neither local nor global synchrony, in region B there is local synchrony but no global synchrony, and in both regions C and D there is both local and global synchrony. We note that although both $\overline{r},R>0$ in both regions C and D, the nature of solutions for $r_\sigma$ are qualitatively different, as will be discussed later. Finally, solid and dashed curves indicate bifurcations between these regions and will be discussed as we proceed with the analysis. In the rest of this section, we will study local synchrony, characterized by the community order parameters $z_\sigma$. We will do this by assuming a given value of the global synchrony order parameter $Z = Re^{i\Psi}$. In the next section, we will study the dynamics of $Z$ using a dimensionality reduction on the global scale. We note here that in the rest of the figures in this paper, since we are interested in networks with a large number of communities and a large number of oscillators per communities, we will compare the results from direct numerical simulation of Eq.~(\ref{eqModel}) on networks with large $N_\sigma$ and $C$ with the theoretical curves obtained from our analysis of the continuum limit. 

\begin{figure}[t]
\centering
\epsfig{file =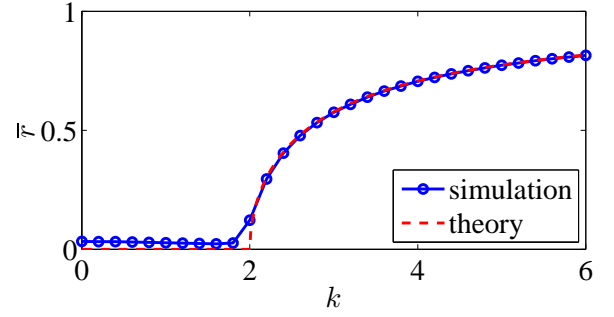, clip =,width=\linewidth }
\caption{(Color online) Average degree of local synchrony $\overline{r}$ versus $k$ from simulation (blue circles) with $C=N_\sigma=400$, $\delta=\Delta=1$ and $K=1$ compared to theoretical prediction in Eq.~(\ref{eqrsigma}) (dashed red).} \label{intra1}
\end{figure}

First we study local synchrony when $R=0$. In this case, from Eqs.~(\ref{eqrho}) and (\ref{eqpsi}) we see that each community decouples from all others and evolves independently. The phase $\psi_\sigma$ of community $\sigma$ moves with velocity $\Omega_\sigma$, and the stable fixed points of Eq.~(\ref{eqrho}) are 
\begin{equation}\label{eqrsigma}
r_\sigma=\left\{\begin{array}{ll} 0 &\text{ if }k-K/C \le 2\delta, \\ \sqrt{1-\frac{2\delta}{k-K/C}} & \text{ otherwise,}\end{array}\right.
\end{equation}
so that all $r_\sigma$ are equal. Bifurcation (i), indicated as a solid black line in Fig.~\ref{BifD}, is described by this analysis, and occurs at $k-K/C=2\delta$. To illustrate this bifurcation, we plot in Fig.~\ref{intra1} the results of simulating the system as $k$ is varied from zero to six with $N_\sigma=C=400$, $\delta=\Delta=1$, and fixed $K=1$ and plot the resulting $\overline{r}$ from simulation (blue circles) against the theoretical prediction of Eq.~(\ref{eqrsigma}) (dashed red). The interpretation of this result is that the oscillators in each community synchronize as in the all-to-all Kuramoto model, but with an effective coupling strength $k-K/C$, which shows that the weak coupling to other independently evolving communities slightly inhibits synchrony.

The analysis above assumes $R=0$. Now, we will analyze local synchrony when $R>0$. In this case some of the communities become synchronized with each other. Given a value of $Z$ (which can be obtained using another dimensionality reduction, as we will show in the next section), community $\sigma$ synchronizes with the mean field [i.e., a solution $\dot{\psi}_\sigma=0,\dot{r}_\sigma=0$ for Eqs.~(\ref{eqrho}) and (\ref{eqpsi}) exists] if 
\begin{equation}\label{eqcommlock}
|\Omega_\sigma|\le KR\frac{r_\sigma^2+1}{2r_\sigma}, 
\end{equation}
in which case 
\begin{equation}\label{eqpsisync}
\psi_\sigma-\Psi=\arcsin\left[\frac{2\Omega_\sigma r_\sigma}{KR(r_\sigma^2+1)}\right],
\end{equation}
and otherwise the community drifts indefinitely. The degree of local synchrony $r_\sigma$ for locked communities can be found by setting $\dot{r}_\sigma$ in Eq.~(\ref{eqrho}) to zero and using Eq.~(\ref{eqpsisync}), which gives the implicit equation
 \begin{align}\label{eqrhoimp}
r_\sigma\delta&=\left(k-\frac{K}{C}\right)r_\sigma\frac{1-r_\sigma^2}{2} \nonumber \\ &\hskip2ex +KR\frac{1-r_\sigma^2}{2}\sqrt{1-\frac{4\Omega_{\sigma}^2r_\sigma^2}{K^2R^2(r_\sigma^2+1)^2}}.
\end{align} 

Eq.~(\ref{eqrhoimp}) determines the steady-state value of $r_{\sigma}$ for locked communities and yields two possible kinds of solutions for $r_\sigma$: either Eq.~(\ref{eqrhoimp}) has a real solution for every $\Omega_\sigma$, or it has a real solution for only some $\Omega_\sigma$. It can be shown that when $k-K/C\le 2\delta$, Eq.~(\ref{eqrhoimp}) has a real solution for all $\Omega_\sigma$, and thus each community becomes phase-locked and each $r_\sigma$ reaches a fixed point as $t\to\infty$. On the other hand, if $k-K/C>2\delta$, there is a real solution for only some $\Omega_\sigma$ with magnitude less than a critical locking frequency, which we denote as $\widetilde{\Omega}$. In this case communities with $|\Omega_\sigma|\le\widetilde{\Omega}$ phase-lock and $r_\sigma$ is given by the solution of Eq.~(\ref{eqrhoimp}), while other communities continue drifting indefinitely. The phase angle $\psi_{\sigma}$ of a drifting community $\sigma$ increases or decreases monotonically and therefore its order parameter $r_{\sigma}$ might be time dependent, according to Eq.~(\ref{eqrho}). However, assuming a stationary global order parameter with constant $R$ and $\Psi$ (as will be discussed in the next section), the solution of the two-dimensional autonomous system in Eqs.~(\ref{eqrho}) and (\ref{eqpsi}) must approach a limit cycle (this can be shown, for example, using the Poincare-Bendixson theorem \cite{Verhulst1}). To estimate the time averaged value of $r_{\sigma}$ in this limit cycle, we neglect the effect of the cosine term in Eq.~(\ref{eqrho}) over one period and find that the time averaged order parameter for drifting communities is approximated by $\langle r_{\sigma} \rangle = \sqrt{1 - \frac{2\delta}{k-K/C}}$. This value agrees with the solution of Eq. (\ref{eqrhoimp}) when $\Omega_{\sigma}$ is the locking frequency in Eq.~(\ref{eqcommlock}). Therefore, the community locking frequency can be determined by inserting the expression for $\langle r_{\sigma} \rangle$ above into Eq.~(\ref{eqcommlock}), obtaining that communities lock when 
their frequency $\Omega_{\sigma}$ satisfies
\begin{equation}\label{Y} 
|\Omega_{\sigma}| \leq \widetilde{\Omega} = KR\left(1-\frac{\delta^2}{(k-\frac{K}{C}-\delta)^2}\right)^{-1/2}
\end{equation}

\begin{figure}[t]
\centering
\epsfig{file =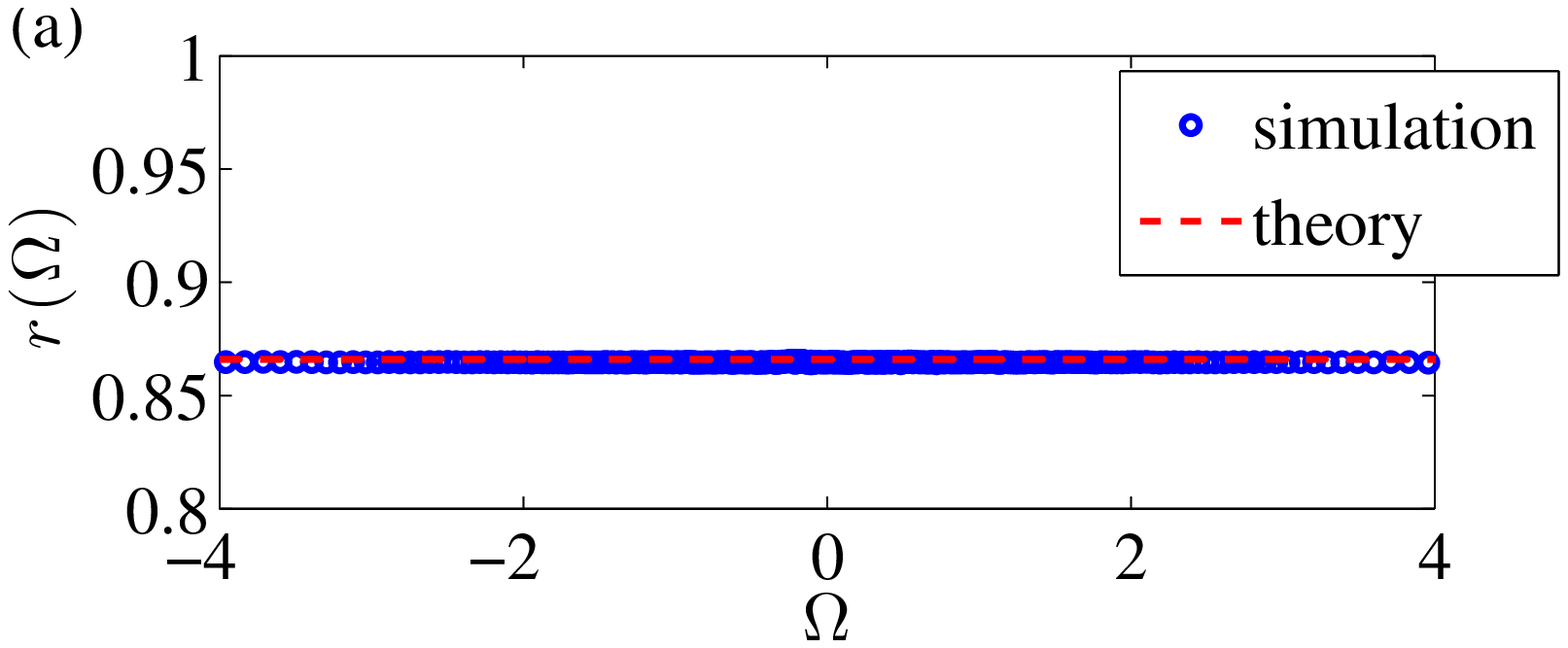, clip =,width=\linewidth } \\
\epsfig{file =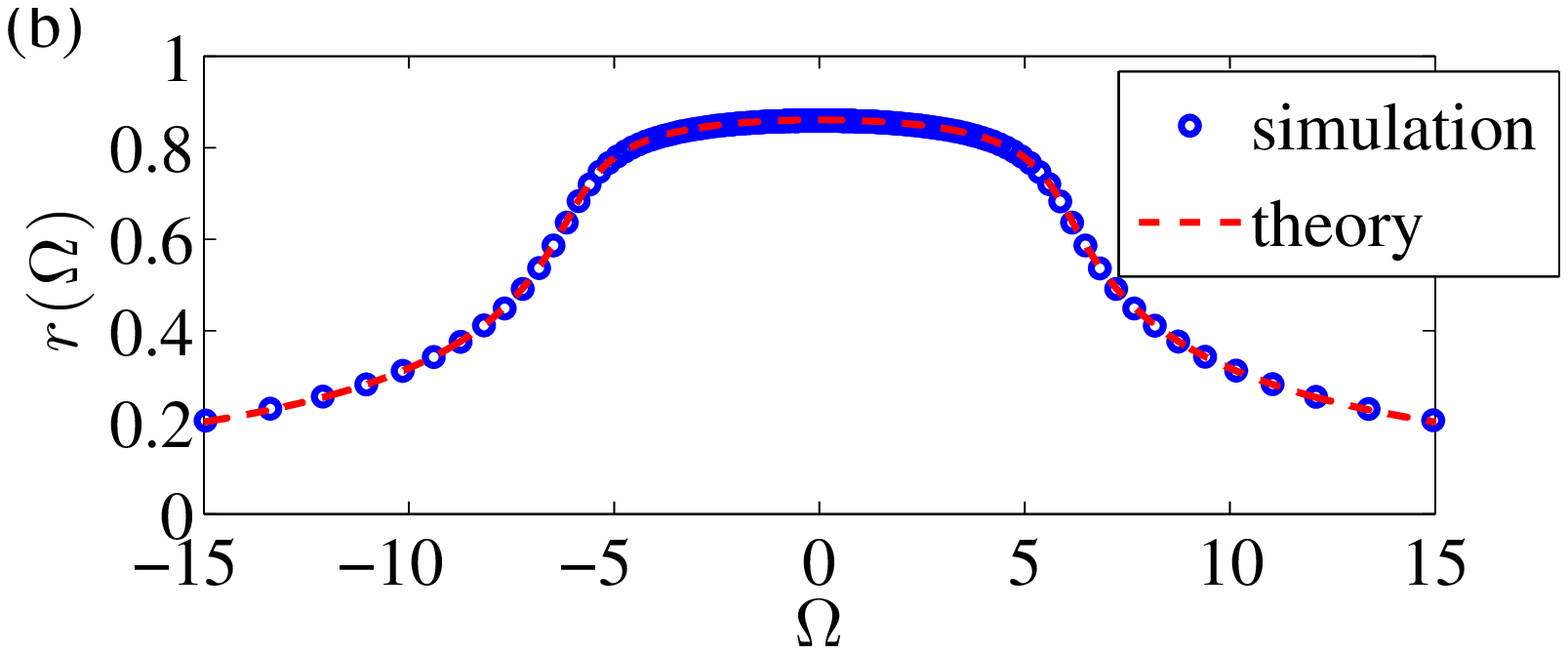, clip =,width=\linewidth } \\
\epsfig{file =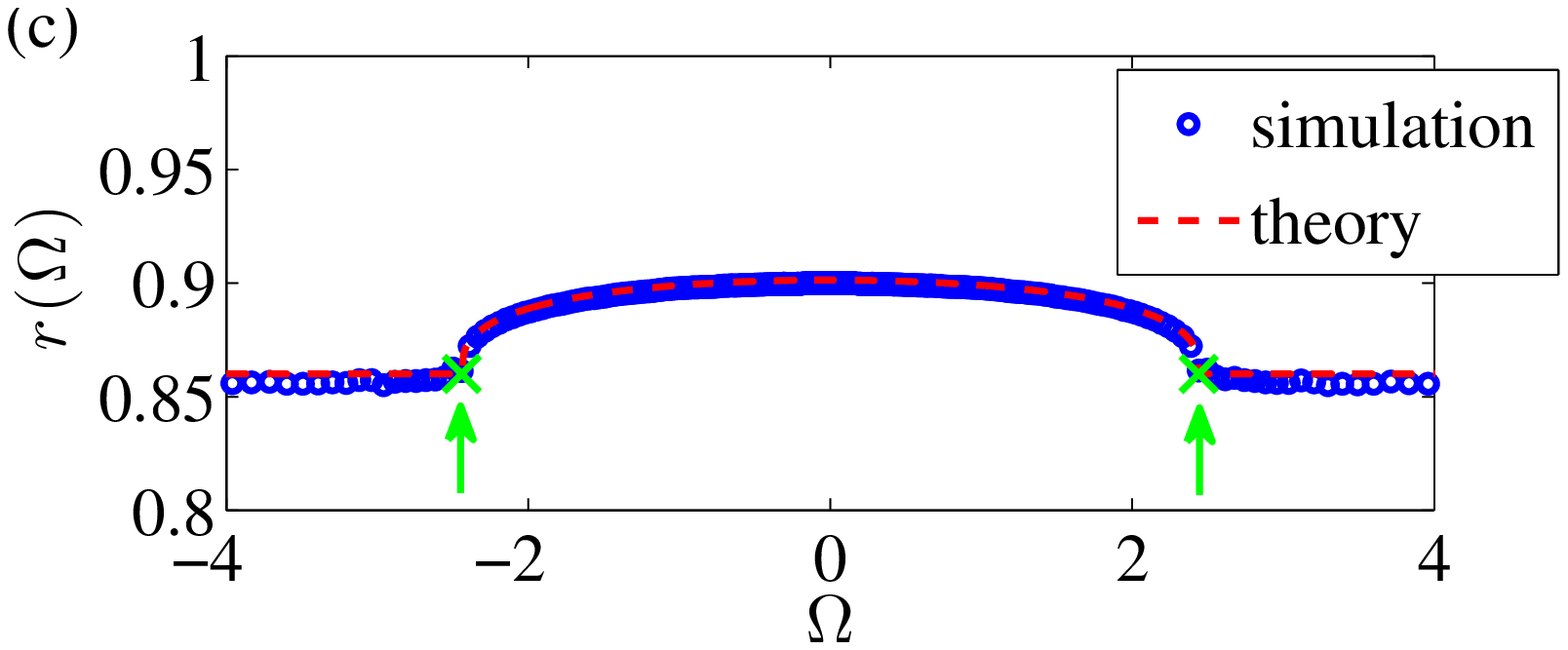, clip =,width=\linewidth }
\caption{(Color online) Time-averaged $r$ vs $\Omega$ from simulation (blue circles) with $C=N_\sigma=400$ and $\delta=\Delta=1$ compared to theoretical prediction (dashed red) for $(K,k)=(1,8)$ (a), $(8,1)$ (b), and $(4,8)$ (c). The vertical arrows indicate the theoretical value for the locking frequency obtained from Eq. (\ref{Y}).} \label{intra2}
\end{figure}

The locking frequency only is defined for $k-K/C > 2\delta$, which defines a new bifurcation. When $k-K/C > 2\delta$ (region D),  the locking frequency $\widetilde{\Omega}$ is finite and only some communities phase-lock. As $k-K/C$ approaches $2\delta$ from above, the locking frequency diverges. For $k-K/C < 2\delta$ (region C), all communities phase lock. The boundary between these two regions for larger $K$ is denoted as bifurcation (ii), and is indicated as a solid black line in Fig. 1. A heuristic interpretation of this transition is that when $k$ is increased through bifurcation (ii), communities with large $|\Omega_{\sigma}|$ desynchronize because the local coupling strength $k$ causes them to prefer an angular velocity $\dot \Psi$ much closer to their own mean frequency $\Omega_{\sigma}$ than the mean frequency of the entire network.

To test Eqs.~(\ref{eqrsigma}), (\ref{eqrhoimp}), and (\ref{Y}), we simulate the system with $N_\sigma=C=400$ and $\delta=\Delta=1$ with $(K,k)=(1,8)$, $(8,1)$, and $(4,8)$ (parameters from regions B, C, and D, respectively) and plot time-averaged $r_\sigma$ as a function of $\Omega_\sigma$ in Figs. \ref{intra2}(a), (b), and (c), respectively. Results from direct simulation are plotted in blue circles and compared to theoretical predictions, which are plotted as dashed red curves. Fig.~\ref{intra2}(a) corresponds to region B, where $R = 0$ and $r_{\sigma}$ is given by Eq. (\ref{eqrsigma}) and is therefore independent of $\sigma$. Fig.~\ref{intra2}(b) corresponds to region C, where all communities lock and their order parameter $r_{\sigma}$ is a solution of Eq.~(\ref{eqrhoimp}). Fig.~\ref{intra2}(c) corresponds to region D, where some communities lock and their order parameter $r_{\sigma}$ is a solution of Eq. (\ref{eqrhoimp}), and other communities drift and their order parameter $r_{\sigma}$ is independent of $\sigma$ and given by $\langle r_\sigma\rangle$. The vertical arrows indicate the theoretical value for the locking frequency obtained from Eq. (\ref{Y}). Theoretical results match very well with the numerical simulations.

\section{Global dimensionality reduction}

In the previous section, we studied local synchrony by assuming a steady-state value for the global synchrony order parameter $Z=Re^{i\Psi}$. We now discuss how the global order parameter can be found by making a second dimensionality reduction on a global scale. As we previously let $N_\sigma$ tend to infinity in order to enter a continuum description within each community, we now consider the limit $C\to\infty$ and introduce the density function $F(\psi,\Omega,r,t)$ that describes the density of communities with average phase $\psi$, mean natural frequency $\Omega$, and degree of local synchrony $r$ at time $t$. In analogy with individual oscillators, the number of communities is conserved and $F$ must satisfy the continuity equation $\partial_tF+\partial_\psi(F\dot{\psi})+\partial_r(F\dot{r})=0$. However, we find that the degrees of local synchrony $r$ quickly reach a stationary distribution, so we seek solutions where $\partial_r(F\dot{r})=0$. In analogy to the classical Kuramoto model, we find that $r_\sigma$ approaches a fixed point if community $\sigma$ phase-locks, or otherwise forms a stationary distribution with other drifting $r$'s. With Eq.~(\ref{eqpsi}) the continuity equation becomes
\begin{equation}\label{eqCont}
\partial_t F + \partial_\psi\left\{F\left[\Omega+K\left(\frac{r^2+1}{2r}\right)\text{Im}(Ze^{-i\psi})\right]\right\}=0,
\end{equation}
where $r=r(\Omega,R)$ is the steady-state value of $r$ given by Eq.~(\ref{eqrsigma}) or implicitly by Eq.~(\ref{eqrhoimp}).

Like Eq.~(\ref{eqContIntra}), Eq.~(\ref{eqCont}) is of the form studied by Ott and Antonsen in Refs.~\cite{OA1,OA2}, and can be solved with a similar ansatz. Thus, we make the ansatz
\begin{align}\label{eqAnsInter}
F(\psi,\Omega,r,t)=\frac{G(\Omega)}{2\pi}\left(1+\sum_{n=1}^\infty A^n(\Omega,r,t)e^{in\psi}+c.c.\right).
\end{align}
Inserting Eq.~(\ref{eqAnsInter}) into Eq.~(\ref{eqCont}), we find that
\begin{align}\label{eqAdot}
\dot{A}+i\Omega A+\frac{K}{4}\left(\frac{r^2+1}{r}\right)(A^2Z-Z^*)=0.
\end{align}

\begin{figure}[t]
\centering
\epsfig{file =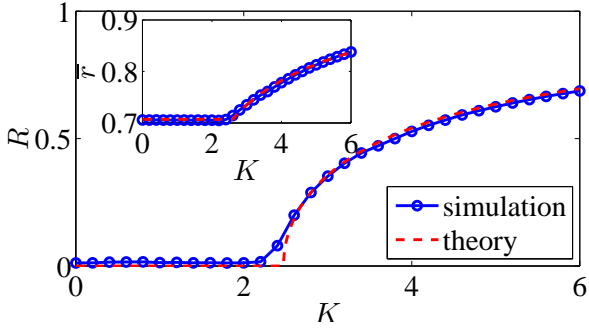, clip =,width=\linewidth } 
\caption{(Color online) Degree of global synchrony $R$ (main) and average local synchrony $\overline{r}$ (inset) versus $K$ from simulation (blue circles) with $N_\sigma=C=400$, $\delta=\Delta=1$, and $k=4$ compared to theoretical prediction from Eqs.~(\ref{eqRss}), (\ref{rhat}), (\ref{eqsolve1}) and (\ref{eqsolve2}) (dashed red).} \label{inter}
\end{figure}

We calculate $Z$ as:
\begin{align}
Z&=\int_{-\infty}^\infty\int_0^{2\pi}F(\psi,\Omega,r,t)re^{i\psi}d\psi d\Omega\nonumber\\
&=\int_{-\infty}^\infty G(\Omega)A^*(\Omega,r,t)r d\Omega.\label{eqZ1}
\end{align}
Since $r(\Omega,R)$ is defined implicitly by Eq.~(\ref{eqrhoimp}) for locked communities and by Eq.~(\ref{eqrsigma}) for any drifting communities, it is potentially piecewise-defined and not smooth. However, to a very good approximation we can do this integral using residues by considering the solution $\widetilde{r}(\Omega,R)$ of Eq.~(\ref{eqrhoimp}) for $|\Omega| < \widetilde{\Omega}$ which is real and positive for $\mbox{Im}(\Omega) \to 0^-$ as a function of complex $\Omega$. The function $\widetilde{r}$ is analytic when $\mbox{Im}(\Omega) < 0$, and its real part converges to $r(\Omega,R)$ as $\mbox{Im}(\Omega) \to 0^-$ with $|\Omega| < \widetilde{\Omega}$, while its imaginary part converges to an odd function. As $\mbox{Im}(\Omega) \to 0^-$ for $|\Omega| > \widetilde{\Omega}$, the real part of $\widetilde{r}$ differs from Eq.~(\ref{eqrsigma}) by a bounded amount. If $G(\Omega)$ decays so quickly that the error in approximating $r$ by $\widetilde{r}$ for $|\Omega| > \widetilde{\Omega}$ can be neglected when computing the integral, we can approximate the integral above by the integral which has $\widetilde{r}$ instead of $r$ (due to the symmetry of $G$, the imaginary part of $\widetilde{r}$ does not contribute to the integral). The integral with $\widetilde{r}$ on the real line can be done by deforming the contour of integration to the line connecting $z_1=-B-i\epsilon$ to $z_2=B-i\epsilon$, where $B$, $\epsilon > 0$, and closing the contour with the semi-circle in the negative complex plane connecting $z_2$ to $z_1$. Using the residue theorem, and taking $B\to\infty$ and $\epsilon \to 0$, we obtain
\begin{align}
 Z\approx\widehat{r}A^*(-i\Delta,\widehat{r},t), \label{eqRA}
\end{align}
where we have defined $\widehat{r}\equiv \widetilde{r}(-i\Delta,R)$. For the Lorenzian distribution with $\Delta = 1$, we expect this approximation to be excellent when $\widetilde{\Omega}\gtrsim 4$, but the agreement between the direct numerical simulation of Eqs.~(\ref{eqModel}) and the theoretical predictions is very good even for situations in which $\widetilde{\Omega}$ is smaller [e.g., Fig.~\ref{hier} (a) close to the transition for $R$]. We note that if $(k,K)$ is in region C [see Fig.~\ref{BifD}] using $\widetilde{r}$ to evaluate the integral in Eq.~(\ref{eqZ1}) is exact since all communities lock and are described by Eq.~(\ref{eqrhoimp}).

Evaluating Eq.~(\ref{eqAdot}) at $\Omega=-i\Delta$ and $r=\widehat{r}$ closes the complex dynamics for $Z$:
\begin{align}
\dot{Z}&+\Delta Z+\frac{K}{4}\left(\widehat{r}^2+1\right)\left(\frac{Z^2}{\widehat{r}^2}Z^*-Z\right)=0.
\end{align}
The evolution of $R$ and $\Psi$ are given by
\begin{align}
\dot{R}&=-\Delta R+\frac{K}{4}R\left(\widehat{r}^2+1\right)\left(1-\frac{R^2}{\widehat{r}^2}\right),\label{eqRho}\\
\dot{\Psi} &=0.\label{eqPsi}
\end{align}
We note that these equations are valid provided that (a) $F$ is in the manifold of Poisson kernels [i.e. is of the form in Eq.~(\ref{eqAnsInter})] and (b) the distribution of degrees of local synchrony $r$ remains stationary as the system evolves. Regarding assumption (a), Ref.~\cite{OA2} shows that in the Kuramoto model all solutions approach this manifold as $t\to\infty$. The stable fixed points of Eq.~(\ref{eqRho}) are 
\begin{equation}\label{eqRss}
R=\left\{\begin{array}{ll}0 & \text{ if }K\le\frac{4\Delta}{\widehat{r}^2+1}, \\ \widehat{r}\sqrt{1-\frac{4\Delta}{K(\widehat{r}^2+1)}} &\text{ otherwise.} \end{array}\right.
\end{equation}

To eliminate $\widehat{r}$ we assume nonzero $R$ (and thus $\widehat{r}\ge\sqrt{4\Delta/K-1}$), and insert Eq.~(\ref{eqRss}) into Eq.~(\ref{eqrhoimp}) with $\Omega_\sigma=-i\Delta$. We choose the real, positive solution given by
\begin{equation}\label{rhat}
\widehat{r}=\sqrt{\frac{\Delta-\delta+\sqrt{(k+K-\delta)^2-2(k+K+\delta)\Delta+\Delta^2}}{k+K}},
\end{equation}
which we insert back into Eq.~(\ref{eqRss}) to obtain $R$. We note that other solutions for $\widehat{r}$ are purely imaginary or negative. From the top line of Eq.~(\ref{eqRss}), the imaginary solutions for $\widehat r$ result in a critical value for $K$ larger than $4\Delta$, while real solutions result in a critical value smaller than $4\Delta$, and thus we choose the positive real solution (the negative solution results in $R< 0$). Finally, to calculate the bifurcation curve for the onset of global synchrony, we let $\widehat{r}\to\sqrt{4\Delta/K-1}^+$ which yields the curve $k=\frac{\delta K}{K-2\Delta}-\frac{K}{2}$. This curve is indicated as a dashed black curve in Fig.~\ref{BifD} and gives bifurcation (iii) from region A to C and bifurcation (iv) from region B to D.

We now seek to compute the mean degree of local synchrony $\overline{r}$. In the large $C$ limit we consider here, $\overline{r}$ is given by an integral equation. If $(K,k)$ is in region C, i.e. $k\le2\delta$, then since each community becomes phase-locked, we simply have
\begin{align}\label{eqsolve1}
\overline{r}=\int_{-\infty}^\infty G(\Omega)r(\Omega,R)d\Omega.
\end{align}
However, if $(K,k)$ is in region D, i.e. $k>2\delta$, then because some communities phase lock and some do not, we have that
\begin{align}\label{eqsolve2}
\overline{r}=\int_{|\Omega|\le \widetilde{\Omega}(R)}G(\Omega)r(\Omega,R)d\Omega + \int_{|\Omega|>\widetilde{\Omega}(R)}G(\Omega)\langle r\rangle d\Omega,
\end{align}
where $\widetilde{\Omega}$ is the locking frequency given by Eq.~(\ref{Y}).

To illustrate these results, we simulate the system with $N_\sigma=C=400$, $\delta=\Delta=1$, $k=4$, and let $K$ vary between zero and six. In Fig.~\ref{inter} we plot $R$ (main) and $\overline{r}$ (inset) from simulation in blue circles and the theoretical predictions from Eqs.~(\ref{eqRss}), (\ref{rhat}), (\ref{eqsolve1}) and (\ref{eqsolve2}) in dashed red. Theoretical predictions agree well with simulations.

\section{Hierarchical Synchrony}

\begin{figure}[t]
\centering
\epsfig{file =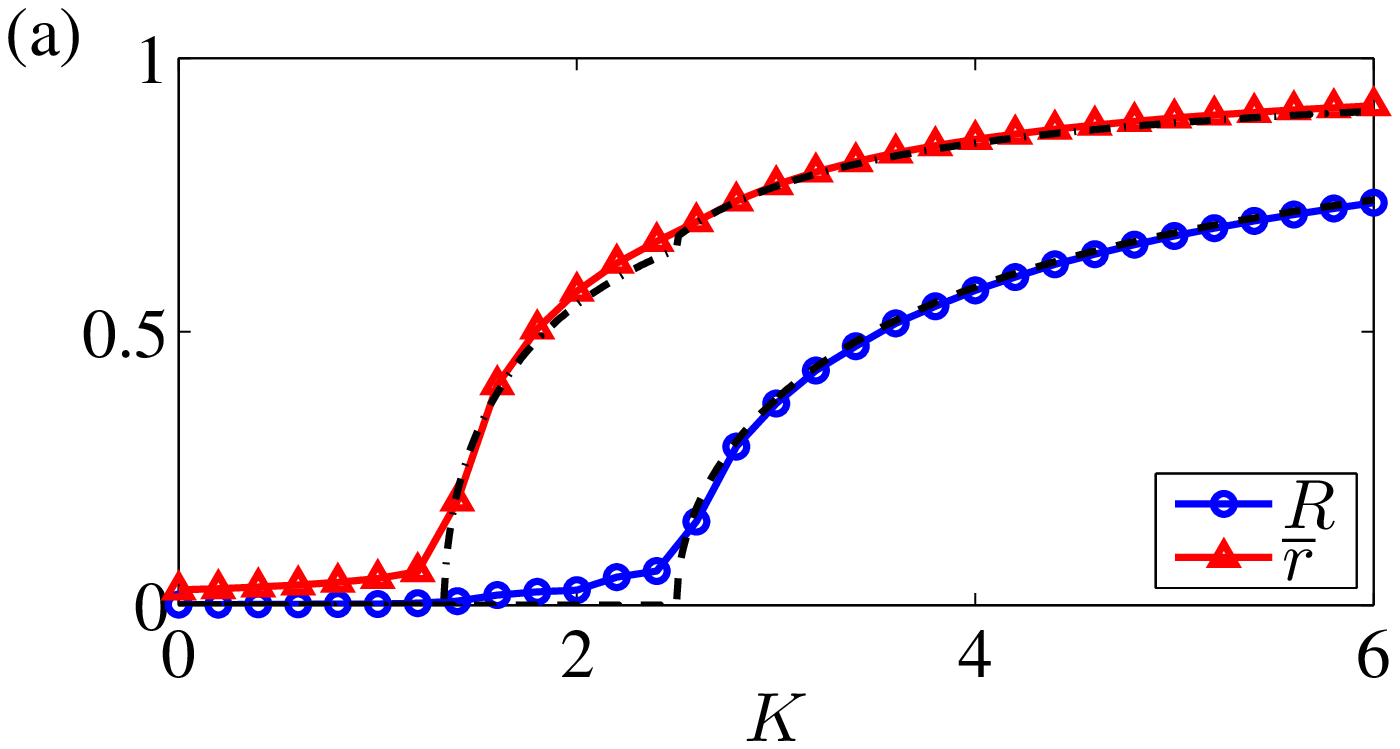, clip =,width=\linewidth } \\
\epsfig{file =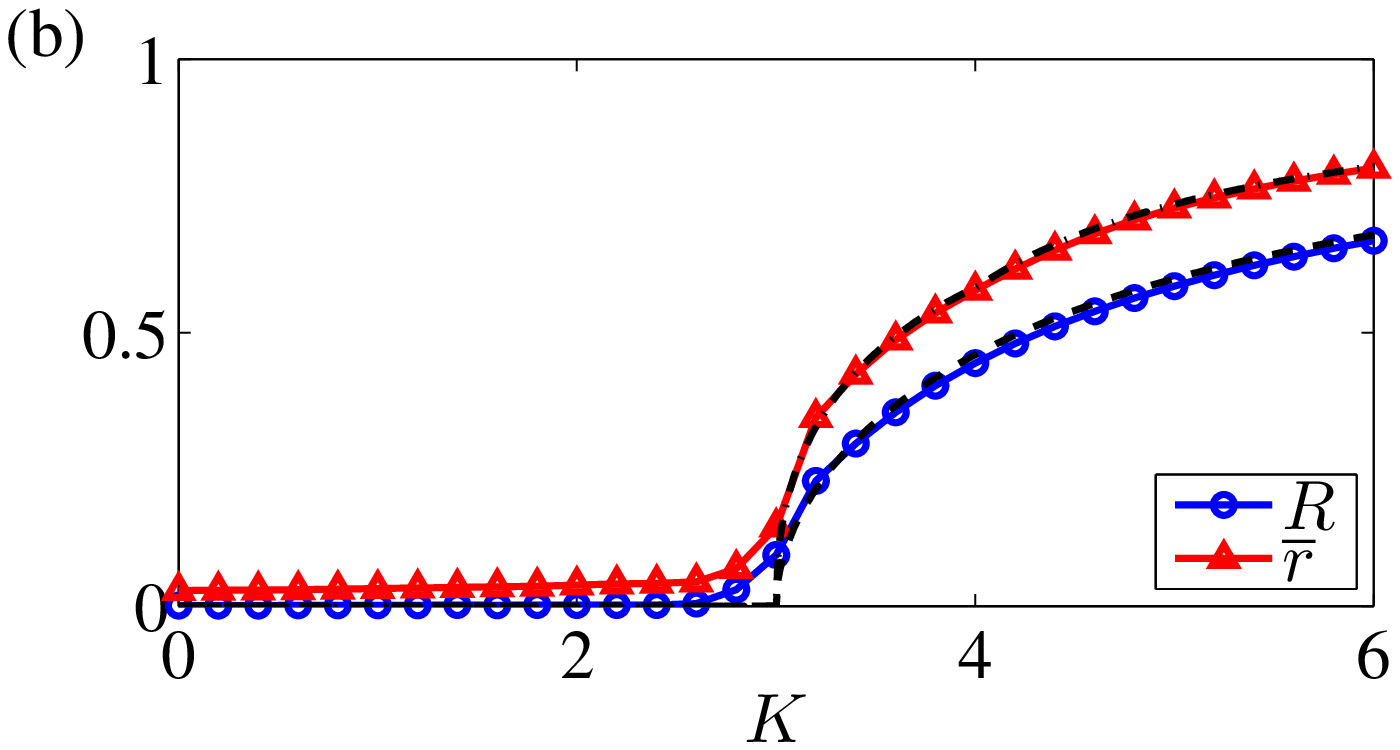, clip =,width=\linewidth }
\caption{(Color online) Degrees of global synchrony $R$ (blue circles) and average local synchrony $\overline{r}$ (red triangle) along paths (a) $k=3K/2$ and (b) $k=K/2$ from simulation with $N_\sigma=C=1000$ and $\delta=\Delta=1$.} \label{hier}
\end{figure}

With a complete understanding of both local and global synchrony in the system studied above, we now discuss hierarchical synchrony. We consider moving slowly (compared with $\Delta^{-1}$) along some path in $(K,k)$ parameter space, restricting paths to lines starting at $(0,0)$ for simplicity. From our analysis we find that bifurcations intersect at $(K,k)=(\Delta-\delta+\sqrt{\Delta^2+\delta^2+6\Delta\delta},2\delta)$. Thus, for lines $k=mK$, if $m>m_c=\frac{2\delta}{\Delta-\delta+\sqrt{\Delta^2+\delta^2+6\Delta\delta}}$ the onset of local synchrony occurs before the onset of global synchrony. On the other hand, if $m<m_c$ the onset of local and global synchrony occur simultaneously. Choosing $m_1=3/2$ and $m_2=1/2$, in Figs.~\ref{hier}~(a) and (b) we plot the steady-state values of $R$ and $\overline{r}$ resulting from moving along the lines $k=m_1K$ and $k=m_2K$, respectively, for $N_\sigma=C=1000$ and $\delta=\Delta=1$. We note that $N_\sigma=C=1000$ is used in these simulations rather than $400$ as in the previous simulations because we find that finite-size effects are more prevalent near bifurcation (iii). This is most likely due to the fact that at this bifurcation the onset of local and global synchrony occurs simultaneously. The values of $R$ and $\overline{r}$ from simulation are plotted in blue circles and red triangles, respectively, with theoretical predictions plotted in black dashed and dot-dashed, respectively. Note that for these parameters $m_c=1/\sqrt{2}$, so $m_1>m_c>m_2$, and accordingly we see a separation of local and global onset in Fig.~\ref{hier}(a), but not in Fig.~\ref{hier}(b).

We interpret these results as follows. Along paths where $k>m_cK$ local coupling effects dominate global coupling effects. In this case the community structure is strong enough to yield a hierarchical ordering of synchrony, i.e. a separation in the onset of local and global synchrony. However, when $k<m_cK$ global coupling effects dominate local coupling effects. In this case the community structure is weak enough to yield a simultaneous onset of local and global synchrony.

\section{Heterogeneities}

We now discuss how the results above generalize when some of the assumptions previously used are relaxed. We allow for heterogeneities in both the sizes of communities and spread in frequency distributions $g_\sigma(\omega)$, i.e. we allow $\eta_\sigma$ and $\delta_\sigma$ to vary from community to community. We also allow the local and global coupling strengths to vary, letting $K^{\sigma\sigma'}=k^\sigma$ for $\sigma=\sigma'$ and $K^\sigma$ for $\sigma\ne\sigma'$.

Beginning with local synchrony, we carry out a dimensionality reduction on the local scale and obtain the following ODEs:
\begin{align}
\dot{r}_\sigma & = -\delta_\sigma r_\sigma  + C\eta_\sigma\left(k^\sigma-\frac{K^\sigma}{C}\right)r_\sigma \frac{1-r_\sigma^2}{2}\nonumber\\ &\hskip2ex+K^\sigma\frac{1-r_\sigma^2}{2} R \cos(\Psi-\psi_\sigma)\label{heterolocal1}\\
\dot{\psi}_\sigma &=\Omega_\sigma +K^\sigma\left(\frac{r_\sigma^2+1}{2 r_\sigma}\right)R\sin(\Psi-\psi_\sigma).\label{heterolocal2}
\end{align}

Thus, when $R=0$, we have that
\begin{equation}
r_\sigma=\left\{\begin{array}{ll} 0 & \text{ if } C\eta_\sigma(k^\sigma-K^\sigma/C)\le2\delta_\sigma\\ \sqrt{1-\frac{2\delta_\sigma}{C\eta_\sigma(k^\sigma-K^\sigma/C)}} & \text{ otherwise.}\end{array}\right.
\end{equation}
The onset of local synchrony in community $\sigma$ occurs at $k^\sigma-K^\sigma/C=2\delta_\sigma/(C\eta_\sigma)$, i.e. in general synchrony occurs at different values for different communities. When $R>0$, community $\sigma$ becomes phase-locked if 
\begin{equation}
|\Omega_\sigma|\le K^\sigma R\frac{r_\sigma^2+1}{2r_\sigma},
\end{equation}
in which case $r_\sigma$ satisfies
\begin{align}
\delta_\sigma r_\sigma&=C\eta_\sigma\left(k^\sigma-\frac{K^\sigma}{C}\right)r_\sigma\frac{1-r_\sigma^2}{2} \nonumber \\ &\hskip2ex +K^\sigma R\frac{1-r_\sigma^2}{2}\sqrt{1-\frac{4\Omega_{\sigma}^2r_\sigma^2}{K^{\sigma2}R^2(r_\sigma^2+1)^2}},
\end{align}
otherwise community $\sigma$ will drift. Note that for a given value of $R$, the behavior of $r_\sigma$ depends not only on $\Omega_\sigma$, but also $C\eta_\sigma$, $\delta_\sigma$, $k^\sigma$, and $K^\sigma$, so in general there is no single locking frequency $\widetilde{\Omega}$ that separates locked and drifting communities at $\Omega_\sigma=\pm\widetilde{\Omega}$.

To study global synchrony, we again perform a dimensionality reduction on the global scale. Since $\eta_\sigma$, $\delta_\sigma$, $k^\sigma$, and $K^\sigma$ vary from community to community, after sending $C\to\infty$ we introduce the density function $F(\psi,\Omega,r,\eta,\delta,k,K,t)$ that represents the fraction of communities with phase $\psi$, mean natural frequency $\Omega$, degree of local synchrony $r$, size $\eta$, frequency distribution spread $\delta$, and local and global coupling strengths $k$ and $K$ at time $t$. Noting that $r_\sigma$ depends on $\eta_\sigma$, $\delta_\sigma$, $k^\sigma$, and $K^\sigma$ and again looking for solutions with stationary $r_\sigma$, $F$ satisfies the continuity equation
\begin{equation}
\partial_t F + \partial_\psi\left\{F\left[\Omega+K\left(\frac{r^2+1}{2r}\right)\text{Im}(Ze^{-i\psi})\right]\right\}=0,
\end{equation}
where now $r$ depends on $\eta$, $\delta$, $k$, and $K$ in addition to $\Omega$ and $R$. We assume that for each community the mean frequency $\Omega_\sigma$, size $\eta_\sigma$, frequency distribution spread $\delta_\sigma$, and local and global coupling strengths $k^\sigma$ and $K^\sigma$ are all chosen independently and make the ansatz
\begin{align}
&F(\psi,\Omega,r,\eta,\delta,k,K,t)=\frac{G(\Omega)H(\eta)D(\delta)J(k)L(K)}{2\pi}\nonumber \\&\hskip4ex\times\left(1+\sum_{n=1}^\infty A^n(\Omega,r,\eta,\delta,k,K,t)e^{in\psi}+c.c.\right),
\end{align}
which yields the ODE 
\begin{align}\label{heteroglobal1}
\dot{A}+i\Omega A+\frac{K}{4}\left(\frac{r^2+1}{r}\right)(A^2Z-Z^*)=0.
\end{align}

Finally in the continuum limit $Z$ can be calculated by the integral
\begin{align}\label{heteroglobal2}
Z&=\int_0^\infty \int_0^\infty \int_0^\infty \int_0^1 \int_{-\infty}^\infty\int_0^{2\pi} F(\psi,\Omega,r,\eta,\delta,k,K,t) \nonumber \\ &\hskip10ex\times re^{i\psi} d\psi d\Omega d\eta d\delta dk dK\nonumber \\
 &=\int_0^\infty \int_0^\infty \int_0^\infty \int_0^1 H(\eta)D(\delta)J(k)L(K)\nonumber \\ &\hskip10ex\times\widehat{r}A^*(-i\Gamma,\widetilde{r},\eta,\delta,k,K,t)d\eta d\delta dk dK.
\end{align}

Equations.~(\ref{heteroglobal1}) and (\ref{heteroglobal2}) govern the global synchrony of the system and must be solved self-consistently with the local dynamics, governed by Eqs.~(\ref{heterolocal1}) and (\ref{heterolocal2}). For arbitrary distribution functions $H(\eta)$, $D(\delta)$, $J(k)$, and $L(K)$ the integral in Eq.~(\ref{heteroglobal2}) might need to be evaluated numerically, but for certain choices, e.g. exponentials or linear combinations of Dirac delta functions, further analytical results are attainable but not presented here.

\section{Discussion}

We have described and solved fully the steady-state dynamics of coupled phase oscillators on a modular network with a large number of oscillators in each community and a large number of communities. In particular, we have studied local and global synchrony, i.e. synchrony within and between communities, respectively. First we assumed a large number of oscillators in each community and used a local dimensionality reduction to study local synchrony. Next, when the number of communities is large, we showed that a global dimensionality reduction can be done to study global synchrony. Our analytical results shed light on the phenomenon of hierarchical synchrony, characterized by synchronization on a local scale before it occurs on a global scale, which occurs when the community structure of the network is strong enough. The system analyzed in this paper modeled synchrony on a network with two community levels, but synchrony on networks with more levels, e.g. communities with subcommunities, can be modeled in a similar way and analogous analytical results can be obtained. 

Although we have assumed strong uniform coupling within communities and weak uniform coupling between communities, we conjecture, based on preliminary numerical experiments, that the system studied in this paper is in some cases a good quantitative model for networks where links between oscillators in the same community are dense and links between oscillators in different communities are sparse. 

An interesting result is that the system of planar oscillators representing community interactions [Eqs.~(\ref{eqrho}) and (\ref{eqpsi})] admits an approximate low dimensional description. The analysis of community synchrony in Sec.~IV is, to the best of our knowledge, the first low-dimensional description of oscillator systems in which each oscillator has a phase and an associated oscillation amplitude. Other systems of coupled planar oscillators could be analyzed in the same way.

\section*{Acknowledgements}
The work of PSS and JGR was supported by NSF Grant No. DMS-0908221.

\bibliographystyle{plain}


\end{document}